\documentclass[12pt]{article}
\usepackage[utf8]{inputenc}
\usepackage{graphicx}
\usepackage[english]{babel}
\usepackage{amsmath}
\usepackage{amssymb}
\usepackage[margin=1in]{geometry}
\usepackage{hyperref}
\usepackage{float}
\usepackage{multirow}
\usepackage{latexsym}
\usepackage{caption}
\usepackage{babelbib}
\usepackage{latexsym}
\usepackage[T1]{fontenc}
\usepackage{tabularx}
\usepackage{dcolumn}
\usepackage{eurosym}
\usepackage{setspace}
\usepackage{color}
\usepackage[usenames,dvipsnames]{xcolor}
\usepackage{geometry}
\usepackage{multicol}
\usepackage{caption}
\usepackage{subfigure}
\usepackage{amsthm}
\usepackage{stmaryrd}
\usepackage{enumitem}
\usepackage{physics}
\usepackage{bbold}
\usepackage[bb=ams]{mathalfa}

\newcommand{\im}[2]{{#1}_{\scriptscriptstyle{#2}}}
\newcommand{\irm}[2]{{#1}_{\scriptscriptstyle{\textrm{#2}}}}

\newcommand{\Los}{M}
\newcommand{\Lof}{m}
\newcommand{\Lcs}{\text{G}}
\newcommand{\Lcf}{\text{KG}}

\newcommand{\oscpb}{(\osv,\osm)}

\newcommand{\osv}{q}
\newcommand{\osvb}{(q)}
\newcommand{\osvz}{\im{\osv}{0}}
\newcommand{\osm}{\im{p}{q}}
\newcommand{\osM}{\im{P}{q}}
\newcommand{\ofv}{x}
\newcommand{\ofm}{y}

\newcommand{\csv}{b}
\newcommand{\csvz}{\im{b}{0}}
\newcommand{\csvb}{(b)}
\newcommand{\csm}{\im{p}{b}}
\newcommand{\cfv}{\im{\phi}{0}}
\newcommand{\cfm}{\im{\pi}{0}}
\newcommand{\cscp}{\csv, \csm}
\newcommand{\cscpb}{(\csv, \csm)}
\newcommand{\qcfv}{\im{\qf{\phi}}{0}}
\newcommand{\qcfm}{\im{\qf{\pi}}{0}}

\newcommand{\Ps}[1]{\im{\Gamma}{#1}}
\newcommand{\Us}[1]{\qs{1}_{\Hs{#1}}}
\newcommand{\Uf}[1]{\qf{1}_{\Hf{#1}}}

\newcommand{\Hs}[1]{\im{\mathcal{H}}{#1}}
\newcommand{\Hf}[1]{\im{\mathcal{H}}{#1}}

\newcommand{\Hfpd}[2]{\im{\mathcal{H}}{#1}^{#2}}

\newcommand{\Hi}{\mathcal{H}}

\newcommand{\osmass}{M}
\newcommand{\ofmass}{m}

\newcommand{\cmg}{\im{m}{\kappa}}
\newcommand{\ml}{\im{m}{\lambda}^{\scriptscriptstyle{(\csv)}}}
\newcommand{\ok}{\im{\omega}{\kappa}}
\newcommand{\ol}{\im{\omega}{0}}
\newcommand{\olam}{\im{\omega}{\lambda}}
\newcommand{\ff}[1]{\omega^{\scriptscriptstyle{(#1)}}}

\newcommand{\lb}[1]{l^{\scriptscriptstyle{(#1)}}}
\newcommand{\f}[1]{f^{\scriptscriptstyle{(#1)}}}

\newcommand{\qs}[1]{\hat{#1}}
\newcommand{\qf}[1]{\pmb{#1}}

\newcommand{\hoscp}{\qf{h}^{\oscpb}}
\newcommand{\h}[2]{\im{\qf{h}}{#1}^{\scriptscriptstyle{#2}}}
\newcommand{\heff}[2]{\im{\qf{h}}{\text{eff}#1}^{\scriptscriptstyle{#2}}}
\newcommand{\ao}[1]{\qf{a}^{\scriptscriptstyle{(#1)}}}
\newcommand{\ad}[1]{\left(\qf{a}^{\scriptscriptstyle{(#1)}}\right)^{\dagger}}

\newcommand{\ef}[2]{\im{e}{#1}^{\scriptscriptstyle{(#2)}}}
\newcommand{\Ef}[1]{\im{E}{#1}}
\newcommand{\Efpd}[2]{\im{E}{#1}^{\scriptscriptstyle{#2}}}

\newcommand{\alp}[2]{\im{\alpha}{#1}^{\scriptscriptstyle{#2}}}

\newcommand{\pipd}[2]{\im{\qf{\pi}}{#1}^{\scriptscriptstyle{#2}}}
\newcommand{\p}{\qf{\pi}}
\newcommand{\prors}[3]{\im{e}{#1}^{\scriptscriptstyle{#3}} \im{\left\langle \im{e}{#1#2}^{\scriptscriptstyle{#3}}, \cdot \right\rangle}{\Lof}} 
\newcommand{\prols}[3]{\im{e}{#1#2}^{\scriptscriptstyle{#3}} \im{\left\langle \im{e}{#1}^{\scriptscriptstyle{#3}}, \cdot \right\rangle}{\Lof}} 
\newcommand{\proru}[4]{\im{e}{#1}^{\scriptscriptstyle{#3}} \im{\left\langle \im{e}{#1#2}^{\scriptscriptstyle{#4}}, \cdot \right\rangle}{\Lof}} 
\newcommand{\prolu}[4]{\im{e}{#1}^{\scriptscriptstyle{#3}} \im{\left\langle \im{e}{#1#2}^{\scriptscriptstyle{#4}}, \cdot \right\rangle}{\Lof}} 

\newcommand{\cprors}[3]{\im{e}{#1}^{\scriptscriptstyle{#3}} \im{\left\langle \im{e}{#1#2}^{\scriptscriptstyle{#3}}, \cdot \right\rangle}{\Lcf}} 
\newcommand{\cprols}[3]{\im{e}{#1#2}^{\scriptscriptstyle{#3}} \im{\left\langle \im{e}{#1}^{\scriptscriptstyle{#3}}, \cdot \right\rangle}{\Lcf}} 
\newcommand{\cproru}[4]{\im{e}{#1}^{\scriptscriptstyle{#3}} \im{\left\langle \im{e}{#1#2}^{\scriptscriptstyle{#4}}, \cdot \right\rangle}{\Lcf}} 
\newcommand{\cprolu}[4]{\im{e}{#1}^{\scriptscriptstyle{#3}} \im{\left\langle \im{e}{#1#2}^{\scriptscriptstyle{#4}}, \cdot \right\rangle}{\Lcf}} 

\newcommand{\us}[2]{\im{\qf{u}}{#1}^{\scriptscriptstyle{#2}}}

\newcommand{\pp}{\varepsilon}

\newcommand{\rel}{n}
\newcommand{\refs}{\mathrm{R}}

\newcommand{\dr}{\mathrm{d}}

\newcommand{\ccr}{canonical commutation relations}
\newcommand{\sapt}{space adiabatic perturbation theory}

\usepackage{enumitem}
\makeatletter
\newcommand{\mylabel}[2]{#2\def\@currentlabel{#2}\label{#1}}
\makeatother

\title{{\sf Quantum Cosmological Backreactions II:}\\
 {\sf Purely Homogeneous Quantum Cosmology}} 
\author{
{\sf J. Neuser}$^1$\thanks{{\sf 
jonas.neuser@fau.de}},
{\sf S. Schander}$^1$\thanks{{\sf 
susanne.schander@gravity.fau.de}},
{\sf T. Thiemann}$^1$\thanks{{\sf 
thomas.thiemann@gravity.fau.de}}\\
\\
{\sf $^1$ Inst. for Quantum Gravity, FAU Erlangen -- N\"urnberg,}\\
{\sf Staudtstr. 7, 91058 Erlangen, Germany}\\
}
\date{{\small\sf \today}}

\begin{document}

\maketitle

{\sf

\begin{abstract}
This is the second paper in a series of four in which we use space adiabatic 
methods in order to incorporate backreactions among the homogeneous and between the homogeneous and inhomogeneous degrees of freedom in quantum cosmological perturbation theory. 

The purpose of the present paper is twofold. On the one hand it illustrates the formalism of space adiabatic perturbation theory (SAPT) for 
two simple quantum mechanical toy models. On the other it proves the main 
point, namely that backreactions lead to additional correction terms in effective Hamiltonians that one would otherwise neglect in a crude Born-Oppenheimer approximation.  

The first model that we consider is a harmonic oscillator coupled to an anharmonic oscillator. We chose it because it displays many similarities with the more interesting 
second model describing the coupling between an inflaton and gravity restricted to the purely homogeneous and isotropic sector. These results have potential phenomenological consequences in particular for quantum cosmological theories describing big bounces such as 
Loop Quantum Cosmology (LQC).
\end{abstract}

\newpage

\tableofcontents

\newpage

\section{Introduction}
\label{s1}

In recent years, significant progress has been made in the field of 
(quantum) cosmology due to major improvements in experimental precision 
\cite{1} and theoretical modelling \cite{2}. One of the most fascinating 
aspects of cosmology is the fact that it allows us to probe primordial 
physics as early as when inflation began \cite{3} or even beyond thus
entering the Planck era \cite{4}. 

It is beyond reasonable doubt that the Planck era must be described by 
a theory of quantum gravity \cite{5} since classical general relativity
predicts that the energy density diverges at the big bang. Unfortunately
today we do not have an accepted quantum gravity theory, in part, because
there is no experimental evidence for it that may help constructing such a 
theory. The problem is the huge Planck mass that suppresses any quantum 
gravity based particle physics effects in earth based experiments 
by sixteen order of magnitude as compared to the typical mass scale of the 
standard model.

Quantum cosmology is therefore perhaps our best chance to detect fingerprints 
of quantum gravity physics that was in action during the Planck era. 
On the one hand the coupling between quantum geometry and quantum matter
together with a quantum to classical transition of fluctuations of the latter 
could lead to indirect observations of the quantum nature of geometry 
by means the density distribution of matter (radiation, baryons, dark matter)
in the universe. On the other, primordial quantum tensor fluctuations 
could directly source primordial gravitational waves. In both cases, 
inflation enhances the amplitude of these originally tiny effects to 
scales that are possibly in the reach of current experimental resolution.

The theoretical processing of these quantum geometrical fluctuations 
into observational data is very challenging and involves a number of stages.
The first stage is the construction of a fundamental theory of quantum 
gravity. The second is an efficient description of the quantum cosmological 
sector within that theory. The third is the direct or indirect 
quantum to classical transition \cite{7} 
of the quantum geometrical fluctuations 
as described above. Finally, the fourth consists in converting these 
classical fluctuations into data that are measured by current experiments.

In this paper we will be concerned with the first and second step. We roughly
follow the idea of the hybrid approach \cite{8} 
to Loop Quantum Cosmology (LQC) \cite{9}: In classical cosmology 
it is meaningful to separate the degrees of freedom into homogeneous and
inhomogemeous ones where the latter are considered as perturbations 
of the former. In principle, any quantity such as the Ricci scalar can 
be expanded into those perturbations. In this way, the non-polynomial 
nature of the Einstein-Hilbert action is kept for the homogeneous degrees of 
freedom, while any finite order approximant is polynomial in the inhomogeneous
perturbations. One then can quantise the homogeneous degrees of freedom 
such that the non-polynomial structure of the corresponding 
operators is taken fully into account e.g. by using LQC methods
while the inhomogeneous degrees of 
freedom can be quantised by using the powerful machinery of quantum fields 
in curved spacetimes (qft in cst) \cite{10}. 

What we aim to add to the current state of the art is to try to capture 
the backreaction between the homogeneous and inhomogeneous degrees of freedom
as precisely as possible. We believe that this is very crucial in order 
to be able to make contact to observational data because the quantum 
fluctuations of all degrees of freedom including the homogeneous ones 
are expected to be very strong and far from 
semiclassical during the Planck era. Thus it is unclear whether 
semiclassical approximations of the fluctuations \cite{11} 
during the Planck era are sufficiently accurate 
especially in vue of the inflationary amplification of these effects.

The challenge lies in the following problem: On the one hand,
in order to use the machinery 
of qft in cst we need as an input a {\it classical metric} on which the quantum 
fields propagate. On the other, that {\it metric is fundamentally quantum}
and in the current situation strongly fluctuating. To reconcile 
these apparently conflicting frameworks we need to somehow describe 
the metric operator in classical terms. The idea is similar to the 
framework of quantum fields in non - commutative spacetimes \cite{12}
where one quantises the spacetime coordinate dependence of the quantum fields 
by using their Fourier transform and replacing the Fourier phase by 
the corresponding Weyl element. It turns out that this procedure is well 
known in the literature \cite{13} and is called space adiabatic quantisation.
When there are different mass scales involved such as the matter - geometry 
mass ratio, then this quantisation
scheme comes automatically accompanied by a corresponding perturbation 
expansion called space adiabatic perturbation theory (SAPT).

To avoid confusion, note that here we have two kinds of expansions: One is 
the perturbative order defined by the degree of polynomials of the 
inhomogeneous degrees of freedom. The other is the adiabatic parameter 
defined for instance by a mass ratio. In \cite{14} SAPT was generalised 
from phase spaces which are vector spaces to more general cotangent bundles 
such as the ones that one encounters in Loop Quantum Gravity (LQG) ore more
generally in non Abelian gauge theories. It was pointed out there that 
in order to make SAPT, designed for quantum systems with finitely many degrees 
of freedom, available to quantum field theories certain conditions have to 
be met. These conditions, specifically certain Hilbert-Schmidt requirements,
are not automatically met in quantum cosmology \cite{14}. To make 
progress, we use the observation of \cite{15} that these kind of conditions 
which also occur in related but different questions within the framework 
of qft in cst have better chances to be satisfied if one passes to better 
suited canonical variables before quantisation. The corresponding 
canonical transformation is also subject to the expansion scheme with respect
to the inhomogeneous versus homogeneous variables.

We have laid out the corresponding general theory in \cite{16} where we 
included a brief and self-contained introduction to both SAPT as well the 
relevant aspects of \cite{15}. In addition to the complications originating
from the infinite number of degrees of freedom we encounter further ones that
arise from the non-polynomial and singular structure of the contributions   
of the inhomogeneous degrees of freedom to the Hamiltonian (constraint)
for which we suggest partial solutions.

In this paper we apply the formalism of \cite{16} to two quantum mechanical 
models. The first model, which to the best of our knowledge was not yet treated 
with the metods of SAPT is a fast harmonic oscillator coupled to a slow 
anharmonic one. This model has the advantage that it keeps the mathematical 
complexity relatively low while incorporating important features of the 
more interesting second model. The second model is general relativity coupled 
to a Klein Gordon field (inflaton) and then truncated to the purely 
homogeneous degrees of freedom. In suitable variables and in the 
presence of a positive (negative) cosmological constant, this model can be 
considered as a fast harmonic oscillator coupled to a slow flipped 
(negative energy) harmonic oscillator, that is, the kinetic (and the potential)
term have a sign opposite to that of a harmonic oscillator. 

The purpose of both models is twofold: On the one hand it illustrates the 
SAPT formalism in a relatively simple and familiar context.
On the other hand we will showcase how the SAPT formalism 
extracts very efficiently an effective Hamiltonian for the slow sector 
while incorporating the interaction with the fast one and displays 
the post-Born-Oppenheimer adiabatic correction terms that arise from the
backreaction between the two kind of degrees of freedom. The scond model 
can also be considered as the purely homogeneous truncation of the 
quantum field theoretic models that we treat in two subsequent papers 
of this series \cite{17,18}.  \\
\\
The architecture of this paper is as follows:\\
\\
In section 2 we introduce the first model and carry out the SAPT formalism.
In addition to extracting the adiabatic corrections, we could in principle
go one 
step further in the spectral analysis and combine the adiabatic 
expansion with the framework of quantum mechanical stationary perturbation
theory \cite{19} as the spectrum of the Hamiltonian is pure point.

In section 3 we consider the second model with the same methods. As the 
spectrum is not pure point when the cosmological constant is positive
one cannot resort to stationary perturbation theory \cite{20} but must use 
independent methods \cite{21} for the spectral analysis of the effective 
Hamiltonian constraint. In the (unphysical) negative cosmological constant 
case we again have a pure point spectrum and thus could in principle
be treated as the first model except that now one is just is interested in 
the kernel of of the operator. Due to the unphysical nature of this case 
we will not go much into detail
but just mention that the pure point nature of the spectrum will lead to 
delicate matching conditions which critically influence the size and 
structure of the kernel. In either case the effective Hamiltonian 
constraint is of a rather singular nature. In contrast to the procedure 
followed in LQC we will refrain from using a non standard representation 
of the gravitational homogeneous degrees of freedom which leads to 
an unseparable Hilbert space. Instead we will stay in the Schr\"odinger
representation but use as a computational tool the new dense and invariant 
operator domain described in \cite{22}. \\

In section four we summarise our findings and conclude.

\section{Oscillator Model}

\subsection{The Hamiltonian}

We apply the space adiabatic perturbation scheme to a quantum system which consists of an anharmonic oscillator with heavy mass $\osmass$, coupled to a harmonic oscillator with lighter mass $\ofmass$ and refer the reader to \cite{16} for an explanation of the scheme and the notation employed throughout this series of papers.

The Hilbert space of the anharmonic oscillator 
subsystem is an $L^2$-space and it will be denoted as $\Hs{\Los}$ or as $L^2(\mathbb{R})$. 
The corresponding position and momentum operator are given as linear self-adjoint operators on a dense domain 
$\irm{\mathcal{D}}{\Los} \subset \Hs{\Los}$, namely $\qs{\osv} \in \irm{\mathcal{L}}{sa}(\irm{\mathcal{D}}{\Los})$ 
and $\qs{\osM} \in \irm{\mathcal{L}}{sa}(\irm{\mathcal{D}}{\Los})$, satisfying the commutation relation,
\begin{equation}
\im{\left[ \qs{\osv}, \qs{\osM} \right]}{\Los} = i \, \Us{\Los}.
\end{equation} 
The Hilbert space associated to the harmonic oscillator subsystem will be denoted as $\Hf{\Lof}$. 
In the given case, one could directly restrict the representation of $\Hf{\Lof}$ to the $L^2$-space 
over the configuration variable $\ofv$ of the harmonic oscillator. However, since the adiabatic 
theory only requires that the Hilbert space $\Hf{\Lof}$ be separable, we keep the notation. 
The position operator $\qf{\ofv} \in \irm{\mathcal{L}}{sa}(\irm{\mathcal{D}}{\Lof})$ 
and the momentum operator $\qf{\ofm} \in \irm{\mathcal{L}}{sa}(\irm{\mathcal{D}}{\Lof})$ 
associated to the harmonic oscillator are defined on a dense domain 
$\irm{\mathcal{D}}{\Lof} \subset \Hf{\Lof}$. They satisfy the commutation relation, 
\begin{equation}
\im{\left[ \qf{\ofv}, \qf{\ofm} \right]}{\Lof} = i\, \Uf{\Lof}.
\end{equation}
The state space of the coupled system is the tensor product, $\Hi:= L^2(\mathbb{R}) \otimes \Hf{\Lof}$, 
and the Hamilton operator, acting linearly on $\Hi$, is defined as,
\begin{equation} \label{eq:Application Hamilton Definition}
\qs{\qf{h}} = \frac{\qs{\osM}^2}{2 \osmass} \otimes \Uf{\Lof}
+ \Us{\Los} \otimes \frac{\qf{\ofm}^2}{2\ofmass} 
+ \frac{1}{2} \ofmass \,\left( \ff{\osv} \right)^2 \otimes \qf{\ofv}^2~~ \in \irm{\mathcal{L}}{sa}(L^2(\mathbb{R}) \otimes \Hf{\Lof})
\end{equation}
We introduce then the adiabatic perturbation parameter, $\pp$, which relates the two mass scales with another, namely $\ofmass = \osmass\cdot\pp^2$. In equation 
\eqref{eq:Application Hamilton Definition}, $\omega: \mathbb{R} \rightarrow \mathbb{R}$ is a measurable function which serves as a $\osv$-dependent frequency in the space adiabatic perturbation picture that we will employ from section \ref{subseq:OM_Parameter Dependent HO} onwards. 
It is given as,
\begin{equation} \label{frequency}
\ff{\osv} = \ol \left(1+ \frac{\osv^2}{L^2} \right).
\end{equation}  
The parameter $L\in \mathbb{R}$ has dimension of length and plays the role of the coupling constant between the two oscillators (the coupling vanishes in the limit
$L\to \infty$). Note that the parametric dependence on the ``heavy'' canonical pair, $\oscpb$, is indicated by upper indices in order to distinguish it from the dependence on the ``light'' canonical pair $(\ofv,\ofm)$. For the full Hamilton operator, we employ the measurable functional calculus for $\omega$ 
in order to obtain $\omega^{(\qs{\osv})}$. \\
In a first step, we ``prepare'' the Hamilton operator, \eqref{eq:Application Hamilton Definition}, 
for the space adiabatic perturbation scheme. Therefore, we substitute $\osmass = \ofmass \cdot \varepsilon^{-2}$, 
and we define a rescaled momentum operator for the anharmonic oscillator, $\im{\qs{p}}{\osv}:= \pp \qs{\osM}$. We can then write,
\begin{equation}
\im{\qs{\qf{h}}}{0} :=\qs{\qf{h}} = \frac{\im{\qs{p}}{\osv}^2}{2 \ofmass} \otimes \Uf{\Lof}
+ \Us{\Los} \otimes \frac{\qf{\ofm}^2}{2 \ofmass} + \frac{1}{2} \ofmass\,\left( \ff{\osv}\right)^2 \otimes \qf{\ofv}^2.
\end{equation}
Even though the Hamilton operator does not exhibit any direct dependence on $\pp$ in this presentation, 
the $\pp$-dependence reappears in the new commutation relation between $\im{\qs{p}}{\osv}$ and $\qs{\osm}$,
\begin{equation}
\im{\left[\qs{\osv},\qs{\osm} \right]}{\Los} = i\, \pp\, \Us{\Los}.
\end{equation}
Then, the section is organised as follows: First we consider $\osv$ as a parameter and 
solve the spectral problem for the part of the Hamiltonian that depends on the fast
degrees of freedom. After that we go step by step through the space adiabatic perturbation scheme in order to construct the effective Hamilton operator to second order.
Therefore, we need the Moyal projector and the Moyal unitarity, both at first order. Finally, we compute the effective Hamiltonian to second order for each energy band defined by the fast oscillator. We highlight the adiabatic correction terms as compared to a crude Born-Oppenheimer approximation. Finally, we comment on the solutions to the effective Hamilton operators.

\subsection{Space Adiabatic Perturbative Scheme} \label{sec:OM_SAPT}

\subsubsection{The Parameter-Dependent Harmonic Oscillator} \label{subseq:OM_Parameter Dependent HO}
We first examine the $\osv$-parameter-dependent eigenvalue problem of the principal symbol 
$\h{0}{\oscpb} \in S_1^2(\Ps, \mathcal{B}(\Hf{\Lof}))$ of the Hamiltonian symbol. This means that the quantum operators of the slow system $(\qs{\osv},\im{\qs{p}}{\osv})$ will formally be considered as simple parameters, $\oscpb \in \mathbb{R}^{2d}$. In the given case, this is already the full Hamiltonian symbol, i.e.,
\begin{equation} \label{eq:OM_h0}
\h{0}{\oscpb} = \frac{\osm^2}{2\ofmass} \Uf{\Los} + \frac{\qf{\ofm}^2}{2 \ofmass} 
+ \frac{1}{2} \ofmass\, \ff{\osv} \qf{\ofv}^2.
\end{equation}
For fixed $\oscpb$, the first term represents a constant zero-point energy which plays no role for the form of the eigenfunctions of $\h{0}{}$. The remaining part 
represents the Hamilton operator of the harmonic oscillator with a $\osv$-dependent frequency $\ff{\osv}$. 
The solutions of the corresponding eigenvalue problem suggest to rephrase the Hamilton operator 
in terms of (an)other pair(s) of unitarily equivalent canonical conjugate operators, namely 
the ``creation''- and ``annihilation''-operators, $\ad{\osv}$ and $\ao{\osv}$. 
These satisfy the \ccr,
\begin{equation}
\im{\left[ \ao{\osv}, \ad{\osv} \right]}{\Lof} = \Uf{\Lof}.
\end{equation}
They relate to the former operators $\qf{\ofm}$ and $\qf{\ofv}$ as,
\begin{equation}
\ao{\osv} = \sqrt{\frac{\ofmass\, \ff{\osv}}{2}} \left( \qf{\ofv}
+ \frac{i}{\ofmass\, \ff{\osv}} \qf{\ofm} \right),~ \ad{\osv} = 
\sqrt{\frac{\ofmass\, \ff{\osv}}{2}} \left( \qf{\ofv} - \frac{i}{\ofmass\, \ff{\osv}} \qf{\ofm} \right).
\end{equation}
The eigenvalue problem of $\h{0}{\oscpb}$ in the position Schrödinger representation, namely,
\begin{equation} \label{eq:OM_EVPinitial}
\h{0}{\oscpb} \ef{n}{\osv}(\ofv)= \Efpd{n}{\oscpb} \ef{n}{\osv}(\ofv)
\end{equation}
has the well-known solutions,
\begin{equation} \label{eq:OM_h0solutions}
\ef{n}{\osv}(\ofv) = \frac{1}{\sqrt{2^n n!}} \cdot \left( \frac{1}{(\lb{\osv})^2 \pi} \right)^{\frac{1}{4}} 
\cdot e^{- \frac{\ofv^2}{2 l^2}} \cdot \im{\mathrm{H}}{n} \left( \frac{\ofv}{\lb{\osv}} \right),
\end{equation}
where $\lb{\osv} = \sqrt{\ofmass\,\ff{\osv}}^{\,-1}$ is the fast oscillator length and $\im{\mathrm{H}}{n}$ 
is the Hermite polynomial of order $n$. In particular, it is,
\begin{equation} \label{eq:Hermite}
\im{\mathrm{H}}{n}(\tilde{x}) = (-1)^n e^{\tilde{x}^2} \frac{\mathrm{d}^n}{d \tilde{x}^n} \left( e^{-\tilde{x}^2} \right).
\end{equation}
The corresponding eigenenergies read,
\begin{equation}
\Efpd{n}{\oscpb} = \frac{\osm^2}{2\ofmass} +  \ff{\osv} \left( n + \frac{1}{2} \right).
\end{equation}
Since the perturbation scheme requires the $q$-derivatives of the eigenspace projectors $\pipd{n}{\osvb}$ associated to the eigensolutions $\ef{n}{\osv}$, 
it is necessary to know the $\osv$-derivatives of the eigenfunctions $\ef{n}{\osv}(\ofv)$. Knowing that all excited states 
relate to the vacuum state, $\ef{0}{\osv}$, by means of,
\begin{equation}
\ef{n}{\osv}(\ofv) = \frac{\left( \ad{\osv} \right)^n}{\sqrt{n!}}\, \ef{0}{\osv}(\ofv),
\end{equation}
the derivatives of a generic eigenfunction follows from the derivative of $\ef{0}{\osv}(\ofv)$ and 
of $\ad{\osv}$.
For the vacuum state, i.e., the solution of \eqref{eq:OM_h0solutions} with $n=0$, and for $\ad{\osv}$, 
the first $\osv$-derivatives read,
\begin{align}
\frac{\partial}{\partial \osv} \ef{0}{\osv}(\ofv) &= \frac{\partial_{\osv} \lb{\osv}}{\sqrt{2} \lb{\osv}} \ef{2}{\osv}(\ofv)
:= \frac{\f{\osv}}{\sqrt{2}} \ef{2}{\osv}(\ofv), \\
\frac{\partial}{\partial \osv} \ad{\osv} &= - \f{\osv} \ao{\osv}.
\end{align}
Then, the first two $\osv$-derivatives of a generic eigenfunction $\ef{n}{\osv}(\ofv)$ are given by,
\begin{equation} \label{eq:OM_ef_firstDER}
\frac{\partial}{\partial \osv} \ef{n}{\osv} := \f{\osv} \cdot \im{a}{1,n} \ef{n-2}{\osv} + \f{\osv} \cdot  \im{a}{2,n} \ef{n+2}{\osv} =: \im{\alpha}{1,n}  \ef{n-2}{\osv} + \im{\alpha}{2,n}  \ef{n+2}{\osv} 
\end{equation}
where we defined,
\begin{align}
\im{a}{1,n} &:= - \frac{\sqrt{n(n-1)}}{2}, ~~~  \im{a}{2,n} :=  \frac{\sqrt{(n+1)(n+2))}}{2}, \\
\im{\alpha}{1,n} &:= \f{\osv} \cdot \im{a}{1,n}, ~~~ \im{\alpha}{2,n} := \f{\osv} \cdot \im{a}{2,n}
\end{align}

\subsubsection{Structural Ingredients} \label{subsec:OM_Structural_Ingredients}
There a three structural ingredients of the model for the theory to be applicable. For further details, we refer the reader to \cite{16}.
\begin{enumerate}
\item The quantum state space of the system decomposes as a tensor product,
\begin{equation}
\mathcal{H} = \Hs{\Los} \otimes \Hf{\Lof}
\end{equation}
and the dynamics in $\Hs{\Los}$ happens on much larger time scales as compared to the dynamics in $\Hf{\Lof}$. Here, the parameter, $\pp := \sqrt{\ofmass/\osmass}$ represents the separation of these time scales.
\item
The deformation scheme needed for space adiabatic perturbation theory is the standard deformation quantization with the Weyl-ordering. Here, the deformation happens by means of the adiabatic parameter $\pp$.
\item The principal symbol of the Hamilton function, $\h{0}{\oscpb}$ possesses a pointwise isolated part of the spectrum $\im{\sigma}{0,\rel}^{\scriptscriptstyle{\oscpb}}$ for some quantum number $\rel$ of the fast system. Namely the spectrum $\im{\sigma}{0,\rel}^{\scriptscriptstyle{\oscpb}}$ is seperated by a finite gap from $\im{\sigma}{0}^{\scriptscriptstyle{\oscpb}} \setminus \im{\sigma}{0,\rel}^{\scriptscriptstyle{\oscpb}}$. Here, we choose one of the eigenspaces with the energy label $\rel$. For the perturbation scheme, it is irrelevant which of the $n \in \mathbb{N}$ is taken. For distinct natural numbers $n \neq m; n, m \in \mathbb{N}$, the energy values of the corresponding bands never cross, 
\begin{align}
\left| \Efpd{n}{\oscpb}- \Efpd{m}{\oscpb} \right| &= \ff{\osv}\, \left| n- m \right| \\
&= \ol \left(1+ \frac{\osv^2}{L^2} \right) \left| n-m \right| > 0, ~~\forall \osv \in \mathbb{R}^d.
\end{align}
It becomes obvious why the form of $\ff{\osv}$ was chosen as in \eqref{frequency}. In particular, if zero is an admissable value for $\osv$, $\ff{\osv}$ must contain a constant term, $\ol$.
\end{enumerate}
For a general model, the above assumptions may apply only for a subset of energy band parameters $n$. However, for the present model we can allow all $n\in \mathbb{N}$.\\

The scheme of \sapt ~divides into three steps \cite{13,24},
\begin{enumerate}
\item[1)] Construction of the Moyal projector $\pipd{\rel}{\oscpb} \in S^{\infty}(\pp; \Ps{\Los}, \mathcal{B}(\Hf{\Lof}))$.
\item[2)] Construction of the Moyal unitary $\us{\rel}{\oscpb} \in S^{\infty}(\pp; \Ps{\Los}, \mathcal{L}(\Hf{\Lof}))$.
\item[3)] Construction of the effective Hamiltonian $\heff{\rel}{\oscpb}  \in S^{\infty}(\pp; \Ps{\Los}, \mathcal{L}(\Hf{\Lof}))$.
\end{enumerate}
The explicit construction of the above symbols and its relevance will be explained more in detail in the following.
\subsubsection{Construction of the Moyal Projector $\pipd{\rel}{\oscpb}$} \label{subsec:OM_Constr_pi}
The space adiabatic theorem \cite{24} states that there exist subspaces $\im{\qs{\pmb{\Pi}}}{n} \mathcal{H}$ of the full Hilbert space $\Hi$ which are approximately invariant under the time evolution generated by the full Hamiltonian $\qs{\qf{h}}$. Every such subspace is associated with a spectral band $\im{\sigma}{0,\rel}^{\oscpb}$ of the principal symbol $\h{0}{\oscpb}$ of the full Hamiltonian. \\

The projector $\im{\qs{\qf{\Pi}}}{n} \in \mathcal{L}(\mathcal{H})$ is $\im{\mathcal{O}}{0}(\pp^{\infty})$-close to the Weyl quantization of a semiclassical symbol,
\begin{equation} \label{eq:pisemiclsymbol}
\im{\qf{\pi}}{n}^{\oscpb} \asymp \sum_{N \geq 0} \pp^N \im{\qf{\pi}}{n,N}^{\oscpb}, ~~ \im{\qf{\pi}}{n,N}^{\oscpb} \in S^{\infty}(\Ps{\Los}, \mathcal{L}(\Hf{\Lof}))
\end{equation}
i.e., $\im{\qs{\qf{\Pi}}}{n} = \im{\qs{\qf{\pi}}}{n} + \im{\mathcal{O}}{0}(\pp^{\infty})$. This means, that for all $m \in \mathbb{N}$, there exists a constant $\im{C}{m}$ such that $\Vert \im{\qs{\qf{\Pi}}}{n} - \im{\qs{\qf{\pi}}}{n} \Vert_{\mathcal{L}(\mathcal{H})} \leq \im{C}{m} \pp^m$. Thus, for small $\pp$, the true projector $\im{\qs{\qf{\Pi}}}{n}$ and the Weyl quantization $\im{\qs{\qf{\pi}}}{n}$ are very close in norm. The ``$\asymp$'' -- symbol in \eqref{eq:pisemiclsymbol} indicates that $\im{\qf{\pi}}{n}$, which belongs to a certain class of semiclassical symbols for which e.g., the Moyal product is well-defined, is $\pp^N$-close (with respect to the Fréchet semi-norms) to the first $N$ contributions of the formal power series on the right hand side. We say that $\im{\qf{\pi}}{n}$ is a resummation of the given series. \\

The advantage of \sapt~is its iterative character with respect to the perturbation parameter for the construction of the Moyal semiclassical symbols,  \cite{24}. Similar to standard perturbation theory, it provides a cut-off in the quantum equations of motions such that only a \emph{finite} number of fast subspaces $\im{\{ \ef{n}{\osv} \}}{n \in \mathbb{N}}$ are involved at every order in the perturbation series. Thus, the scheme starts at zeroth order with a ``principal'' symbol $\im{\qf{\pi}}{n,0}^{\oscpb} \in S^{\infty}(\Ps{\Los}, \mathcal{L}(\Hf{\Lof}))$. One can exploit that we already have the eigenprojectors $\pipd{n,0}{\osvb}$ at our disposal for every $\oscpb \in \Ps{\Los}$ on some particular subspace of $\Hfpd{\Lof,\rel}{\oscpb}$ associated to the energy eigenvalues $\Efpd{n}{\oscpb}$ of the initial eigenvalue problem \eqref{eq:OM_EVPinitial}. This gives a convenient start for the adiabatic perturbation theory, since a restriction on these eigenprojections corresponds to the full adiabatic limit. There, the fast degrees of freedom adjust instantaneously to the slow motion of the slow degrees of freedom and remain in the same eigenspace. We emphasize that the Weyl quantized operators $\im{\qs{\qf{\pi}}}{n,0}$ in $\Hi$ do \emph{not} commute with $\qs{\qf{h}}$ and are thus not invariant under the evolution generated by the full Hamiltonian. However, the relevant commutator yields only terms in first or higher orders of $\pp$.\\

In order to start the perturbative evaluation, we recall that according to the initial eigenvalue problem, there exists at every point $\oscpb \in \Ps{\Los}$ some eigenspace associated to the quantum number $\rel$ of the fast subsystem. The analysis starts by choosing one such quantum number $\rel$ and the corresponding spectral projection $\pipd{\rel,0}{\osvb}$ of $\hoscp$. Note that this projector exists for every $\oscpb \in \Ps{\Los}$ due to the constant global energy gap. Due to the continuity of the map, $\oscpb \mapsto \hoscp$ also the map, $\oscpb \mapsto \pipd{\rel,0}{\osvb}$, is continuous. Since we restrict the analysis to this particular subspace, we omit the index $\rel$ in the following. \\

In the Schrödinger representation, using the eigenfunctions \eqref{eq:OM_h0solutions}, the projector acts on a wavefunction $\psi(\ofv) \in \Hf{\Lof}$ by,
\begin{equation} \label{eq:pizerodef}
\left[ \pipd{0}{\osvb} \psi \right] (\ofv) = \ef{n}{\osv}(\ofv) \int_{\mathbb{R}} \dr \ofv'\, \ef{n}{\osv}(\ofv') \cdot \psi(\ofv'),
\end{equation}
and $\pipd{0}{\osvb} \in S^{\infty}(\Ps{\Los}, \mathcal{L}(\Hf{\Lof}))$. Recall that the upper parameters ``$\osv$'' indicate the dependence of the eigenvalue problem on the slow variables, while the function parameter ``$\ofv$'' is the wavefunction variable for the fast degrees of freedom. A simpler notation with a generic inner product will be used in the following, namely,
\begin{equation} \label{eq:pi0}
\pipd{0}{\osvb} := \ef{\rel}{\osv} \cdot \im{\left\langle \ef{\rel}{\osv}, \cdot \right\rangle}{\Lof},
\end{equation}
where $\im{\left\langle \cdot, \cdot \right\rangle}{\Lof}: \Hf{\Lof} \times \Hf{\Lof} \rightarrow \mathbb{C}$ denotes the inner product in $\Hf{\Lof}$. \\

The aim of the subsequent analysis is to construct a projector $\pipd{}{\oscpb}$ as a formal symbol series with respect to the deformation parameter $\pp$, i.e., 
\begin{equation} \label{eq:piformalseries}
\pipd{}{\oscpb} = \sum_{N\geq0} \pp^N \pipd{N}{\oscpb},~~~ \pipd{N}{\oscpb} \in S^{\infty}(\mathcal{B}(\Hf{\Lof})),
\end{equation}
with $\pipd{0}{\oscpb}$ chosen as above, \eqref{eq:pizerodef}. The principal symbol of $\h{}{}$ is just $\h{0}{}$ like in \eqref{eq:OM_h0}. The projector $\qs{\qf{\Pi}}$ of the full quantum problem satisfies, in accordance with the space adiabatic theorem,
\begin{equation}
\left[ \qs{\qf{h}}, \qs{\qf{\Pi}} \right] = \im{\mathcal{O}}{0}(\pp^{\infty}).
\end{equation}

Thus, the Weyl quantization of the semiclassical symbol $\pipd{}{\oscpb}$ should also commute $\im{\mathcal{O}}{0}(\pp^{\infty})$-approximately with $\qs{\qf{h}}$. Consequently, $\qs{\p}$ must also be an almost projection in the former sense. Instead of examining the commuations in the operator scheme, deformation quantization suggests to pull them back to phase space by means of the symbolic calculus. The latter employs the star-product ``$\im{\star}{\pp}$'' on the space of semiclassical symbols $S^{\infty}(\pp;\Ps{\Los}, \mathcal{L}(\Hf{\Lof}))$ which is the pull back of the operator product on $\mathcal{H}$ to the Poisson algebra of phase space functions with values in the space of linear operators on $\Hf{\Lof}$, i.e., on $C^{\infty}(\Ps{\Los}, \mathcal{L}(\Hf{\Lof}))$. For two semiclassical symbols $f, g \in S^{\infty}(\pp;\Ps{\Los}, \mathcal{L}(\Hf{\Lof}))$, it is given by,
\begin{equation} \label{eq:starproduct}
f\,\im{\star}{\pp}\,g = \sum_{m,n=0}^{\infty} (-1)^m \left(\frac{i \pp}{2} \right)^{n+m} \frac{1}{m!\,n!} \left(\partial_{\osv}^n \partial_{\osm}^m\,f \right) \left(\partial_{\osv}^m \partial_{\osm}^n\,g \right) \, \in S^{\infty}(\pp;\Ps{\Los}, \mathcal{L}(\Hf{\Lof})).
\end{equation} 

The above requirements for $\qs{\p}$ then translate into,
\begin{enumerate}[label=\textnormal{\arabic*)}]
\item $\p\, \im{\star}{\pp}\, \p = \p$, \label{itm:1}
\item $\p^{\ast} = \p$, \label{itm:2}
\item $\im{\left[ \h{0}{},\p \right]}{\im{\star}{\pp}} = 0$. \label{itm:3}
\end{enumerate}
As shown in \cite{13,24}, the constructed symbol $\p$ satisfying these rules, is unique if there is a gap in the spectrum of $\h{0}{}$ for the chosen subspace.  We develop the construction of $\p$ order by order and we define $\pipd{(k)}{}$ to be the properly constructed symbol up to order $k$, i.e., $\pipd{(M)}{} = \sum_{N= 0}^M \pp^N \pipd{N}{}$. Then, for the construction of $\pipd{N}{}$, we use the above conditions, restricted to the $N$-th order, namely,
\begin{enumerate}[label=\textnormal{\arabic*)}]
\item $\pipd{(N)}{}\,\im{\star}{\pp}\,\pipd{(N)}{}- \pipd{(N)}{}  = \mathcal{O}(\pp^{N+1})$, 
\item $\left(\pipd{(N)}{}\right)^{\ast} - \pipd{(N)}{} = \mathcal{O}(\pp^{N+1})$, 
\item $\im{[\h{}{}, \pipd{(N)}{} ]}{\im{\star}{\pp}} =  \mathcal{O}(\pp^{N+1})$,
\end{enumerate}
where ``$\ast$'' denotes the adjoint of the respective operator. For our purposes, i.e., for computing an effective Hamilton symbol up to second order, it will be sufficient to compute the Moyal projector up to first order, namely, $\pipd{(1)}{} = \pipd{0}{} + \pp \cdot \pipd{1}{}$, for which $\pipd{0}{}$ is already known.\\

Taking $\pipd{0}{}$ as a starting point for the iteration and following the rules \ref{itm:1} to \ref{itm:3}, it is straightforward to compute $\pipd{1}{}$. Here, we start with condition \ref{itm:1} which serves for the construction of the diagonal part $\pipd{1}{\mathrm{D}} := \pipd{0}{}\cdot \pipd{1}{} \cdot\pipd{0}{} + (\Uf{\Lof} - \pipd{0}{})\cdot \pipd{1}{} \cdot  (\Uf{\Lof} - \pipd{0}{}) $ for the first order Moyal projector. The dot product stands for the operator product with respect to the fast degrees of freedom. Equation $1)$ restricted to first order is trivially satisfied, since $\pipd{0}{\osvb}$ is an orthogonal projector in $\Hfpd{\Lof}{\oscpb}$, i.e. $\pipd{0}{} \cdot \pipd{0}{} - \pipd{0}{} =0$. At first order in $\pp$, equation \ref{itm:1} gives \textit{a priori} a non-trivial condition for $\pipd{1}{}$. The latter includes derivatives of $\pipd{0}{}$ with respect to the slow canonical pair $\oscpb$, according to the definition of the star product, \eqref{eq:starproduct}. Note however, that $\pipd{0}{}$ does not depend on the slow momentum, $\osm$, such that the Poisson bracket contributions in the condition vanish. The only non - vanishing terms are thus,
\begin{equation}
\pipd{1}{} \cdot \pipd{0}{} + \pipd{0}{} \cdot \pipd{1}{} - \pipd{1}{} = 0. 
\end{equation}
Indeed, it follows that,
\begin{equation} 
\pipd{1}{\text{D}} = 0.
\end{equation}
The condition \ref{itm:3} gives a result for the off - diagonal part, $\pipd{(1)}{\text{OD}} := \pipd{(1)}{} - \pipd{(1)}{\text{D}}= \pipd{(1)}{}$ of $\pipd{(1)}{}$. Again, at zeroth order in $\pp$ the equation is satisfied a priori, i.e., $\im{\left[ \h{0}{}, \pipd{0}{} \right]}{\Lof} =0$. At first order, condition \ref{itm:3} gives,
\begin{equation} \label{eq:Condpifirst}
\h{0}{}\,\cdot\, \pipd{1}{} - \pipd{1}{}\, \cdot \h{0}{} + \frac{i}{2}\, \im{\left\lbrace \h{0}{}, \pipd{0}{} \right\rbrace}{\Los} - \frac{i}{2}\, \im{\left\lbrace \pipd{0}{}, \h{0}{} \right\rbrace}{\Los} = 0.
\end{equation} 
We multiply $\pipd{0}{}$ from the left and its orthogonal complement $(\Uf{\Lof} - \pipd{0}{})$ from the right in  equation \eqref{eq:Condpifirst}. This restrict the defining equation to the lower off-diagonal part, $\pipd{1}{\text{OD},1} := \pipd{0}{}\cdot \pipd{1}{}\cdot(\Uf{\Lof} - \pipd{0}{})$. We could do the same the other way round which would lead to the upper off-diagonal part, $\pipd{1}{\text{OD},2} := (\Uf{\Lof} - \pipd{0}{})\cdot \pipd{1}{}\cdot\pipd{0}{}$, of $\pipd{1}{}$. Taking both contributions, this gives,
\begin{align}
\pipd{1}{} = \frac{i}{2} &\left( (\h{0}{} - \Ef{\rel}\, \Uf{\Lof})^{-1}\cdot (\Uf{\Lof} - \pipd{0}{}) \cdot\im{\left\lbrace \h{0}{} + \Ef{\rel}\,\Uf{\Lof}, \pipd{0}{} \right\rbrace}{\Los} \cdot \pipd{0}{} \right. \nonumber \\
&~~\left.+\, \pipd{0}{} \cdot\im{\left\lbrace \pipd{0}{}, \h{0}{} + \Ef{\rel} \Uf{\Lof}\right\rbrace}{\Los}\cdot (\h{0}{} - \Ef{\rel}\,\Uf{\Lof})^{-1}\cdot (\Uf{\Lof} - \pipd{0}{}) \right). \label{eq:OM_ProjectorSymbol1_1}
\end{align}
Since furthermore, the eigenstates solely depend on $\osv$, the Poisson bracket simplifies and we get,
\begin{equation}
\pipd{1}{} = i \left(\frac{\partial \Ef{\rel}}{\partial \osm}\right) \left( (\h{0}{} - \Ef{\rel}\,\Uf{\Lof})^{-1}\cdot \left(\frac{\partial \pipd{0}{}}{\partial \osv}\right)\cdot \pipd{0}{} + \pipd{0}{} \cdot\left(\frac{\partial \pipd{0}{}}{\partial \osv}\right) \cdot(\h{0}{} - \Ef{\rel} \Uf{\Lof})^{-1} \right),
\end{equation}
where the second term acts only outside of the relevant subspace as also indicated in \eqref{eq:OM_ProjectorSymbol1_1}. Note as well, that the above computation only makes sense if the analysis is restricted to one particular subspace with one initial energy eigenfunction $\Efpd{n}{\oscpb}$. The inverse of the operator $(\h{0}{} - \Ef{\rel}\,\Uf{\Lof})$ reduces to the a factor,
\begin{equation}
\left( \Efpd{\rel}{\oscpb} - \Efpd{\rel \pm 2}{\oscpb} \right)^{-1} = \left(\mp\, 2\, \ff{\osv}\right)^{-1}.
\end{equation} 
By defining the minimal energy gap as $\im{\Delta}{E} := \ol$ and by using the derivatives, \eqref{eq:OM_ef_firstDER}, we obtain in an explicit presentation,
\begin{equation} \label{eq:pi1}
\pipd{1}{}\!=\!\left(\frac{\partial\Ef{\rel}}{\partial \osm} \right)\! \frac{i}{2 \im{\Delta}{E}}\!\left( \alp{1,\rel}{\osvb} \left( \prors{\rel}{-2}{\osvb} - \prols{\rel}{-2}{\osvb} \right) + \alp{2,\rel}{\osvb} \left(  \prols{\rel}{+2}{\osvb} - \prors{\rel}{+2}{\osvb} \right) \right)\!.
\end{equation}
We note that the corresponding Weyl quantized operator $\qs{\pipd{(1)}{}}$ almost-projects on a subspace of $\Hi$ which depends non-trivially on $\pp$ and the description of the dynamics within this subspace would turn out to be very complicated as explained in \cite{16}. \\
In order to overcome this problem, we construct a unitary symbol $\us{}{} \in S^{\infty}(\pp; \Ps{\Los}, \mathcal{L}(\Hf{\Lof}))$ which maps the dynamical subspace related to $\p$, or more precisely here to $\pipd{(1)}{}$, to a suitable reference subspace $\im{\Hi}{0,n}$. This subspace $\im{\Hi}{0,n} \subset \im{\Hi}{0} = \Hf{\Lof}$ does not depend on $\pp$ and we emphasize that the choice of this subspace is not unique. It is just an auxiliary structure, see \cite{16} for more details and how these energy dependent structures are related to the spectral theorem for the Hamiltonian.

\subsubsection{Construction of the Moyal Unitary $\im{\qf{u}}{\rel}^{\oscpb}$} \label{subsec:Constr_MUnitary_OM}
For the given model, the simplest and physically most convenient choices for a reference subspace $\mathcal{R}_{\rel}$, are associated to the chosen subspace $\Hfpd{\Lof}{(\osvz)}$ for some fixed value $\osvz$ of $\osv$. Without loss of generality, we choose $\osvz \equiv 0$ and we denote it as $\Hf{0}$. The corresponding reference projection is,
\begin{equation}
\pipd{\refs}{} := \ef{\rel}{\osvz} \im{\left\langle  \ef{\rel}{\osvz}, \cdot \right\rangle}{\Lof}. 
\end{equation}
In order to mediate between $\Hfpd{\Lof}{\osvb}$ and $\Hf{0}$, and vice versa, a unitary operator $\us{}{}$ is necessary. The condition of unitarity and the requirement that $\us{0}{}$ should map $\pipd{0}{}$ to $\pipd{\refs}{}$ gives at least the following conditions on $\us{0}{}$,
\begin{enumerate}[label=\textnormal{\arabic*)}]
\item $\us{0}{}\,\cdot\,\pipd{0}{}\,\cdot(\us{0}{})^{\ast} = \pipd{\refs}{}$,
\item $\us{0}{}\,\cdot (\us{0}{})^{\ast} = \Uf{0}$,
\item $(\us{0}{})^{\ast} \cdot \us{0}{} = \Uf{\Lof}$.
\end{enumerate}
Therefore, we employ for $\us{0}{}$ the following operator-valued symbol,
\begin{equation}
\us{0}{\osvb} = \sum_{j \geq 0} \ef{j}{\osvz} \im{\left\langle \ef{j}{\osv}, \cdot \right\rangle}{\Lof}. 
\end{equation}
This choice trivially satisfies the conditions on $\us{0}{}$ and is simple and evident.\\

Taking $\us{0}{}$ and the above conditions as a starting point, we aim to construct iteratively a semiclassical symbol $\us{\rel}{\oscpb} \in S^{\infty}(\pp; \Ps{\Los}, \mathcal{L}(\Hf{\Lof}))$. The formal power series has the form,
\begin{equation}
\us{\rel}{\oscpb} \asymp \sum_{N \geq 0} \pp^N \us{\rel,N}{\oscpb}, ~~~ \us{\rel,N}{\oscpb} \in S^{\infty}(\mathcal{B}(\Hf{\Lof})).
\end{equation}
Transcription of the conditions for $\us{0}{}$ using the $\im{\star}{\pp}$-product, gives for the semiclassical symbol $\us{}{} \in S^{\infty}(\pp; \Ps{\Los},\mathcal{L}(\Hf{\Lof}))$,
\begin{enumerate}[label=\textnormal{\arabic*)}]
\item $\us{}{}\; \im{\star}{\pp}\; \p\; \im{\star}{\pp}\; \left(\us{}{}\right)^{\ast} = \pipd{\refs}{}$
\item $\us{}{}\;\im{\star}{\pp}\; (\us{}{})^{\ast} = \Uf{0}$,
\item $(\us{0}{})^{\ast}\;\im{\star}{\pp}\; \us{}{} = \Uf{\Lof}$.
\end{enumerate}
Like for the Moyal projector, the perturbative equations for $\us{}{}$ read, when considered order by order in $\pp$,
\begin{enumerate}[label=\textnormal{\arabic*)}]
\item $\us{(N)}{}~ \im{\star}{\pp}~ \p~ \im{\star}{\pp}~ \left(\us{(N)}{}\right)^{\ast}- \pipd{\refs}{} = \mathcal{O}(\pp^{N+1})$,
\item $\us{(N)}{}~\im{\star}{\pp}~ \left(\us{(N)}{}\right)^{\ast} -\Uf{\Lof} =  \mathcal{O}(\pp^{N+1})$,
\item $\left(\us{(N)}{}\right)^{\ast}~ \im{\star}{\pp}~ \us{(N)}{} - \Uf{\Lof} =  \mathcal{O}(\pp^{N+1})$.
\end{enumerate}
We restrict the formal power series to first order, i.e., we evaluate the former conditions for $\us{(1)}{\oscpb} = \us{0}{\osvb} + \pp\,\us{1}{\oscpb}$. Given $\us{0}{}$, it is straightforward to show that the hermitian part of $\us{1}{}$ vanishes because $\us{0}{}$ is independent of $\osm$. However, the remaining anti-hermitian contribution is given by,
\begin{equation} \label{eq:Unitary_Firstorder_OM}
\us{1}{} = \us{0}{}\cdot \im{\left[ \pipd{0}{}, \pipd{1}{} \right]}{\Lof},
\end{equation}
and simply inserting equation \eqref{eq:pi1}, we obtain,
\begin{equation}
\us{1}{}\!=\!\left(\frac{\partial \Ef{\rel}}{\partial \osm} \right)\! \frac{i}{2 \im{\Delta}{E}}\!\left(\alp{1,\rel}{\osvb} \left(  \proru{\rel}{-2}{(\osvz)}{\osvb} + \prolu{\rel}{-2}{(\osvz)}{\osvb} \right) - \alp{2,\rel}{\osvb} \left( \proru{\rel}{+2}{(\osvz)}{\osvb} + \prolu{\rel}{+2}{(\osvz)}{\osvb} \right)\right)\!.
\end{equation}

\subsubsection{Construction of the Effective Hamilton Symbol $\heff{,\rel}{\oscpb}$}
The last step of the perturbation scheme consists in pulling the dynamics of the chosen subspace to the $\pp$-independent subspace, $\Hf{0}= \im{\qs{\qf{\Pi}}}{\refs} \Hi$. This essentially means that by applying the unitary operator $\qs{\qf{U}} = \qs{\us{}{}} + \im{\mathcal{O}}{0}(\pp^{\infty})$ on the Hamiltonian $\qs{\h{}{}}$, the action of the latter on elements in $\qs{\qf{\Pi}} \Hi$ is rotated to $\Hf{0}$.  We denote the semiclassical symbol,
\begin{equation} \label{eq:hsymbol_equation}
\heff{}{} \asymp\, \us{}{} ~ \im{\star}{\pp}~ \h{}{} ~ \im{\star}{\pp}~ \left(\us{}{}\right)^{\ast},
\end{equation} 
as the effective Hamiltonian.
Then, the Weyl quantization, $\irm{\qs{\qf{h}}}{eff}$, of the symbol $\heff{}{} \in S^{\infty}(\pp;\Ps{\Los}, \mathcal{L}(\Hf{\Lof}))$ is essentially self-adjoint on the Schwartz space $\mathcal{S}(\mathbb{R}, \Hf{\Lof})$. And in particular, the dynamics of $\qs{\h{}{}}$ are mapped unitarily to $\Hf{0}$, such that,
\begin{align}
\left[ \im{\qs{\qf{h}}}{\mathrm{eff}}, \im{\qs{\qf{\pi}}}{\refs} \right] &=0, \\
e^{-i \qs{\qf{H}}s} - \left(\qs{\qf{u}}\right)^{\ast} e^{-i\,\irm{\qs{\qf{h}}}{eff}\,s}\, \qs{\qf{u}} &= \im{\mathcal{O}}{0}(\pp^{\infty} \left| s \right|), \label{eq:Effh_Almost_Invariance}
\end{align}
where $s\in\mathbb{R}$ is a real (timelike) parameter.\\

We construct $\heff{}{}$ perturbatively by means of equation \eqref{eq:hsymbol_equation} up to second order. We assume for the generic form of the semiclassical symbol $\heff{}{}$,
\begin{equation}
\heff{}{\oscpb} = \sum_{N \geq 0}^2 \pp^N\, \heff{,N}{\oscpb},~~ \heff{,N}{\oscpb} \in S^{\infty}(\Ps{\Los}, \mathcal{L}(\Hf{\Lof}))
\end{equation}
Its restriction up to the $N$-th order, $\heff{,(N)}{}$ is defined as,
\begin{equation} \label{eq:OM_heffCondk}
\heff{,(N)}{} = \us{(N)}{}\,\im{\star}{\pp}\,\h{(N)}{}\,\im{\star}{\pp}\, \left(\us{(N)}{}\right)^{\ast} + \im{\mathcal{O}}{0}(\pp^{N+1}).
\end{equation} 
At zeroth order, this simply gives,
\begin{equation}
\heff{,0}{} = \us{0}{}\,\cdot\,\h{0}{}\,\cdot\,\left(\us{0}{}\right)^{\ast} = \sum_{n\geq 0} \Efpd{n}{\oscpb}\, \ef{0}{\osvz} \im{\left\langle \ef{0}{\osvz}, \cdot \right\rangle}{\Lof}.
\end{equation}
We restrict our attention to the dynamics within the subspace $\Hf{0}$. Thus, we construct the effective Hamilton symbol by applying the projection symbol $\pipd{\refs}{}$ from the left and the right on $\heff{,0}{}$. We label the corresponding symbols with the quantum number $\rel$, i.e., $\heff{,\rel,j}{}$ for $j \in \lbrace 0,1,2 \rbrace$.\\
 
At zeroth order, the effective Hamilton symbol within the relevant subspace is simply given by,
\begin{align} \label{eq:OM_ZeroHeff}
\heff{,\rel,0}{} &= \pipd{\refs}{} \cdot \us{0}{} \cdot \h{0}{} \cdot \left(\us{0}{}\right)^{\ast} \cdot \pipd{\refs}{} =  \Efpd{\rel}{\oscpb}\, \ef{\rel}{\osvz} \im{\left\langle \ef{\rel}{\osvz}, \cdot \right\rangle}{\Lof}\\
&= \left( \frac{\osm^2}{2 m} + \ff{\osv}  \left( \rel + \frac{1}{2} \right) \right) \pipd{\refs}{}.
\end{align}
It is now easy to evaluate the action of this symbol on some generic tensor product wave function in $\Hi = \Hs{\Los} \otimes \Hf{\Lof}$. In particular, the operator associated to the fast subsystem, $\pipd{\refs}{}$ has the eigenfunctions $\ef{n}{\osvz}$, which is the same for every $\oscpb \in \Ps{\Los}$. Thus, one can simply examine the action of the $\oscpb$- dependent function $\Efpd{n}{\oscpb}$ on elements of $\Hs{\Los}$.\\
The Weyl quantization scheme in the Schrödinger representation translates ``$\osv$'' into a multiplication operator, while ``$\osm$'' becomes the derivative operator ``$i\,\pp\,\partial_{\osv}$. The Schrödinger equation for some generic wave function $\Psi \in \Hs{\Los}$ is given by,
\begin{equation}
\left( - \frac{\partial_{\osv}^2}{2 \osmass} + \frac{1}{2}\osmass \im{\tilde{\omega}}{\rel}^2 \osv^2 \right) \im{\Psi}{d,\rel}(\osv) = \im{\tilde{E}}{d,\rel}\im{\Psi}{d,\rel}(\osv),
\end{equation}
where we defined,
\begin{equation}
\im{\tilde{\omega}}{\rel} = \sqrt{\frac{2 \ol}{M L^2}\left(\rel + \frac{1}{2} \right)}, ~~ \im{\tilde{E}}{d,\rel} = \im{E}{d,\rel} - \ol \left(\rel + \frac{1}{2} \right),
\end{equation}
and $E$ is the energy of the full system. \\

This is the Schrödinger equation of a harmonic oscillator with mass parameter $\osmass$ and frequency $\im{\tilde{\omega}}{\rel}$. The eigenfunctions are associated to discrete eigenenergies which are not only labeled by the former quantum number $\rel$ of the fast subsystem, but also by the slow quantum number $d$. The eigenenergies are thus given by,
\begin{equation}
\im{E}{d,\rel} = \ol \left(\rel + \frac{1}{2} \right) + \sqrt{\frac{2 \ol}{M L^2}\left(\rel + \frac{1}{2} \right)} \cdot \left(d+ \frac{1}{2} \right).
\end{equation}
The eigenfunctions are given by,
\begin{equation}
\im{\Psi}{d,\rel}(\osv) = \frac{1}{\sqrt{2^d} d!} \left( \frac{M\,\im{\tilde{\omega}}{\rel}}{\pi} \right)^{\frac{1}{2}} \cdot e^{-\frac{M \im{\tilde{\omega}}{\rel}}{2} \osv^2} \cdot \im{\mathrm{H}}{d} \left(q \sqrt{M\,\im{\tilde{\omega}}{\rel}} \right),
\end{equation}
with the Hermite polynomials $\im{\mathrm{H}}{d}$. The result corresponds to the Born-Oppenheimer solution. In this simplified scheme, the slow degrees of freedom encounter an external potential given by a single energy level of the fast degrees of freedom. It represents the adiabatic limit in which the fast degrees of freedom are constrained to stay within one energy band.\\

For the first order contributions, using \eqref{eq:OM_heffCondk}, we can deduce that,
\begin{equation}
\us{}{}\, \im{\star}{\pp}\, \h{}{} - \heff{,0}{} \, \im{\star}{\pp}\, \us{}{} = \pp\,\h{1}{}\,\im{\star}{\pp}\,\us{}{} + \mathcal{O}(\pp^2) = \pp\,\heff{,1}{}\cdot \us{0}{}+ \mathcal{O}(\pp^2). 
\end{equation}
We employ that the total Hamilton symbol has only a zeroth order contribution, i.e., $\h{}{} = \h{0}{}$, which then yields,
\begin{equation}
\heff{,1}{} = \left(\us{1}{}\cdot \h{0}{} - \heff{,0}{}\cdot \us{1}{} + \frac{i}{2} \im{\left\lbrace \us{0}{}, \h{0}{} \right\rbrace}{\Los} - \frac{i}{2} \im{\left\lbrace \heff{,0}{}, \us{0}{} \right\rbrace}{\Los} \right)\cdot \left(\us{0}{} \right)^{\ast}.
\end{equation} 
Knowing that $\us{1}{}$ has no diagonal contributions and that $\us{0}{}$ does not depend on $\osm$, this condition implies that $\heff{,1}{}$ has no diagonal contributions. Hence, the restriction to the chosen subspace with quantum number $n$ vanishes,
\begin{equation}
\heff{,\rel,1}{} = \frac{i}{2}\, \pipd{\refs}{}\cdot \im{\left\lbrace \us{0}{}, \h{0}{} + \Ef{\rel}\,\Uf{\Lof}\right\rbrace}{\Los}\cdot\left(\us{0}{}\right)^{\ast}\cdot \pipd{\refs}{} =0.
\end{equation}
Note however, that $\heff{,1}{}$ does \emph{not} vanish because of the off-diagonal contributions.\\

The determining equation for the second order contribution of $\heff{}{}$ follows as well from \eqref{eq:OM_heffCondk}. The computations are straightforward, but lenghty, so that we restrict directly to the contributions of the dynamics within the relevant subspace, namely,
\begin{equation} \label{eq:OM_heff_second}
\heff{,\rel,2}{} = \frac{i}{2}\, \pipd{\refs}{} \im{\left\lbrace \us{1}{}, \h{0}{} + \Ef{\rel} \right\rbrace}{\Los} \left(\us{0}{}\right)^{\ast}\, \pipd{\refs}{} - \pipd{\refs}{} \heff{1}{} \us{1}{} \left(\us{0}{} \right)^{\ast}\, \pipd{\refs}{} - \frac{i}{2}  \pipd{\refs}{} \im{\left\lbrace \heff{1}{}, \us{0}{} \right\rbrace}{\Los} \left(\us{0}{}\right)^{\ast}\, \pipd{\refs}{},
\end{equation}
where now the non-vanishing contributions of $\heff{,1}{}$ must be taken into account. Though, their off-diagonal contributions vanish as well. The result is,
\begin{equation}
\heff{,\rel,2}{} =  \frac{L^2}{2\ofmass\,\ol} \left( -\frac{\osm^2 \osv^2}{\ofmass \left(L^2+\osv^2\right)^3} \left(\rel + \frac{1}{2} \right) + \frac{\osv^2}{\left(L^2+\osv^2 \right)^2} \frac{\ol}{2 L^2} \left(\rel^2+\rel+1 \right) \right) \pipd{\refs}{}.
\end{equation}
This prooves our statement that besides the trivial Born-Oppenheimer approximation, further quantum backreaction effects arise for the heavy subsystem. The first step is to evaluate the action of the Weyl quantized Hamilton symbol $\heff{,\rel,(2)}{}$ on elements of $\Hs{\Los}$. \\

We leave the spectral analysis of the operator for another paper.

\section{Cosmological Model}

\subsection{The Hamiltonian}
We consider general relativity with metric $g_{\mu\nu}$ including a cosmological constant $\Lambda$ coupled 
to a Klein-Gordon field $\cfv$ with mass $\im{m}{\Lcf}$. We pull the Einstein-Hilbert and the Klein-Gordon action back to 
the homogeneous and isotropic sector. For our illustrative purposes, we consider the 
manifold $\mathbb{R}\times \mathbb{T}^3$ with torus coordinate length $L$ and flat spatial slices, $k=0$. The pull back is done by the embedding,
\begin{equation}
\dr s^2=-\dr t^2+a(t)^2\; \dr\vec{x}^2,\;\;\phi(t,\vec{x})=\cfv(t)
\end{equation}
with the usual scale factor $a$. The action of the homogeneous model then reduces to,
\begin{equation}
\irm{S}{cos} = L^3 \int_{\mathbb{R}} \dr t\, \left( -\frac{1}{2\,\kappa} \left(6\,\dot{a}^2 a 
+ 2\,\Lambda\,a^3\right) + \frac{1}{2\,\lambda} a^3 \left(\im{\dot{\phi}}{0}^2 - \im{m}{\Lcf}^2 \cfm^2 \right) \right).
\end{equation}
where the integration over the volume of the torus produces a factor $L^3$. Here 
$\kappa$ and $\lambda$ are respectively the coupling constants of general relativity and the Klein-Gordon system,  where $\kappa=8 \pi G$ and $G$ is Newton's constant.

If both $(g_{\mu\nu}, \cfv)$ are dimension free as we assume, then both coupling constants have the same dimension. Thus the dimensionless adiabatic parameter,
\begin{equation}
\pp^2 := \frac{\kappa}{\lambda}.
\end{equation}
is the ratio of the mass squared parameters $m^2:=\frac{\hbar}{c^2}\frac{1}{\lambda}$
and $M^2=\frac{\hbar}{c^2}\frac{1}{\kappa}$, where $M$ is the Planck mass. In what 
follows we will assume that $m\ll M$ and thus $\varepsilon\ll 1$ which is certainly the case if $m$ is in the mass range of a typical standard model particle. It transpires
that in the adiabatic language, gravity is the ``slow'' sector and the Klein-Gordon particle
the ``fast'' one. This may seem counter intuitive when one thinks of the Klein-Gordon field as an inflaton candidate and the inflationary phase when $\cfv$ practically freezes (for small $\im{m}{\Lcf}$) while $a$ expands exponentially. However, note that 
the distinction of slow and fast degrees of freedom uses intrinsically a statistical average over the phase space. For instance, when the system under consideration has a true Hamiltonian bounded from below, one uses the equipartition theorem
(see \cite{16} and references therein). In our case we do not have a true Hamiltonian but rather a Hamiltonian constraint so that the equipartition theorem 
does not apply. However, we can use the constraint itself (basically the Friedman equation) to deduce that for the velocites, $u=\dot{\ln(a)}$ and $v=\im{\dot{\phi}}{0}$, it holds that $u^2/\kappa\approx v^2/\lambda$ for small 
$\Lambda/\kappa$ in scalar field kinetic energy dominated parts of the phase space.\\
  
For the space adiabatic scheme, we need to perform a Hamiltonian analysis. We define the momenta of the scale factor $a$ and the Klein-Gordon variable $\cfv$ as,
\begin{equation} \label{eq:DefMomenta}
\im{p}{a} := \frac{\pp}{L^3}\frac{\partial \irm{L}{cos}}{\partial \dot{a}}, ~~~ \cfm
:= \frac{1}{L^3}\frac{\partial \irm{L}{cos}}{\partial \im{\dot{\phi}}{0}}.
\end{equation}
The Poisson brackets of the canonical variables then enter with a factor 
\begin{equation}
\lbrace a, \im{p}{a} \rbrace = \frac{\varepsilon}{L^3},~~~\lbrace \phi, \im{\pi}{\phi} \rbrace = \frac{1}{L^3}.
\end{equation}
This choice for the fundamental Poisson relations assures that we can identify the masses associated to the homogeneous and isotropic degrees of freedom with the mass of the total system of the torus, as pointed out in \cite{16}.\\

Given the definition of the momenta in \eqref{eq:DefMomenta}, the Legendre transformation generates 
the Hamiltonian,
\begin{equation}
\irm{h}{cos} = L^3\,\lambda \left( -\frac{1}{12}\frac{\im{p}{a}^2}{a} + \frac{\Lambda}{\lambda\,\kappa}a^3 
+ \frac{\cfm^2}{2 a^3} +\frac{1}{2\lambda^2} \im{m}{\Lcf}^2 a^3 \cfv^2 \right).
\end{equation}
For notational reasons, we divide the whole constraint by a constant factor $L^3\,\lambda$ 
and we denote the rescaled constraint as $h$. For simplifying the analysis by means of space 
adiabatic perturbation theory in the following, we switch to ``triad like'' canonical variables 
\begin{equation}
\csv := \pm \sqrt{a^3}, ~~\csm :=\frac{2}{3} \frac{\im{p}{a}}{\sqrt{a}},
\end{equation}
which is a double cover of the original phase space. Note that the range of $b$ consists of two branches, a positive and a negative one. We do not 
restrict to any of them.\\

In order to keep the notation as simple as possible, we introduce the following parameters and functions,
\begin{equation}
\cmg := \frac{8}{3}, ~~ \ok^2 := \frac{3\,\Lambda}{4\,\lambda\,\kappa}, 
~~ \ml := b^2, ~~ \olam^2 := \frac{\im{m}{\Lcf}^2}{\lambda^2}.
\end{equation}
Inserting these definitions into the Hamilton constraint together with the new canonical variables gives, 
\begin{equation}
h = - \frac{\csm^2}{2\,\cmg} + \frac{1}{2} \cmg \ok^2 \csv^2 
+ \frac{\cfm^2}{2 \ml} + \frac{1}{2} \ml \olam^2 \cfv^2.
\end{equation}
The Hilbert space representation of the canonical commutation relations, 
\begin{equation}
\im{\left[ \qs{\csv},\im{\qs{p}}{\csv} \right]}{\Lcs} = i\,\pp\,\Us{\Lcs}, ~~ \im{\left[ \qcfv, \qcfm \right]}{\Lcf} = i\,\pp\,\Uf{\Lcf},
\end{equation}
will be chosen as the tensor product $\Hs{\Lcs} \otimes \Hs{\Lcf}$, where both factors are simply $L_2$-spaces over the real axis with Lebesgue measure $\dr\, \csv$ and $\dr\, \cfv$ respectively. Note that this is not the representation chosen in LQC \cite{9}  for which one motivation is that inverse powers of $a$ or $\csv$ can be made well defined following the technique introduced for LQG \cite{23}. That technique does not work in the presently chosen Schr\"odinger representation. However, one can still find a dense and invariant domain \cite{22} for the Hamiltonian constraint operator of the full system which is sufficient to perform the spectral analysis. In what follows, we perform a systematic step by step space adiabatic perturbation theory treatment of this symbol.

\subsection{Space Adiabatic Perturbation Scheme}
\subsubsection{The Parameter-Dependent Harmonic Oscillator} \label{subseq:HO}
We examine the characteristics of the Hamilton symbol,  
\begin{equation} \label{eq:FullHamiltonSymbol}
\h{}{\cscpb}=  \left(- \frac{\csm^2}{2 \cmg} + 
\frac{1}{2} \cmg \ok^2 \csv^2 \right) \Uf{\Lcf}
+ \frac{\qcfv^2}{2 \ml}  + \frac{1}{2} \ml \olam^2\, \qcfm^2,
\end{equation}
in order to prepare the perturbative spectral analysis by means of space adiabatic perturbation theory. Again, the first term of $\h{}{\cscpb}$ represents a parameter-dependent zero-point energy for the Klein-Gordon quantum system. The second term resembles the Hamilton operator of a harmonic oscillator for the Klein-Gordon system with a $\csv$-dependent mass $\ml$. As for the oscillator model, we stick to this picture of viewing $\csv$ and $\csm$ 
as simple parameters for the moment. The corresponding creation and annihilation operators, $\ao{\csv}$ and $\ad{\csv}$, which depend here parametrically on $\csv$ satisfy the \ccr,
\begin{equation}
\im{\left[\ao{\csv}, \ad{\csv} \right]}{\Lcf} = \Uf{\Lcf}.
\end{equation}
They relate to the former operators $\qcfv$ and $\qcfm$ according to,
\begin{equation}
\ao{\csv} = \sqrt{\frac{\ml \olam}{2}} \left( \qcfv
+ \frac{i}{\ml \olam} \qcfm \right),~~ \ad{\csv} 
= \sqrt{\frac{\ml \olam}{2}} \left( \qcfv 
- \frac{i}{\ml \olam} \qcfm \right).
\end{equation}
The eigenvalue problem of $\h{}{\cscpb}$ for the Klein-Gordon subsector is given in the Schrödinger representation by, 
\begin{equation} \label{eq:EVPinitial}
\h{}{\cscpb} \ef{n}{\csv}(\cfv) = \ef{n}{\cscp}\,\ef{n}{\csv}(\cfv).
\end{equation}
with solutions,
\begin{equation}
\ef{n}{\csv}(\cfv) = \frac{1}{\sqrt{2^n n!}} \cdot \left( \frac{1}{\left(\lb{\csv} \right)^2 \pi} \right)^{\frac{1}{4}} 
\cdot e^{- \frac{\cfv^2}{2\lb{\csv} 2}} \cdot \im{\mathrm{H}}{n} \left( \frac{\cfv}{\lb{\csv}} \right),
\end{equation}
where $\lb{\csv} = \left(\sqrt{\ml \olam} \right)^{-1}$ and $\im{\mathrm{H}}{n}$ the Hermite polynomials of order $n$.
The corresponding eigenenergies are given by,
\begin{equation}
\Efpd{n}{\cscpb} = - \frac{\csm^2}{2 \cmg} + \frac{1}{2} \cmg \ok^2 \csv^2 
+ \olam \left( n + \frac{1}{2} \right).
\end{equation}
The energy gap for nearby energy bands is,
\begin{equation}
\left| \Efpd{n}{\cscpb} - \Efpd{n\pm1}{\cscpb}  \right| =  \left| \olam \right|,
\end{equation}
which is constant. We compute the $\csv$-derivatives of the eigenfunctions $\ef{n}{\csv}$. Analogously to the oscillator model, these are given by,
\begin{align}
\frac{\partial}{\partial \csv}\, \ef{0}{\csv} &= \frac{\partial_{\csv}\, \lb{\csv}}{\sqrt{2}\, \lb{\csv}}\,\ef{2}{\csv} 
:= \frac{\f{\csv}}{\sqrt{2}}\,\ef{2}{\csv}, \\
\frac{\partial}{\partial \csv}\,\ad{\csv} &= - \f{\csv}\,\ao{\csv}.
\end{align}
For our model, it is $\f{\csv}= - 1/\csv$. Thereby, the $\csv$-derivative of a generic eigenfunction $\ef{n}{\csv}$ is given by,
\begin{equation} \label{eq:psi_firstDER}
\frac{\partial}{\partial \csv} \ef{n}{\csv} =  \f{\csv}  \cdot \im{a}{1,n} \ef{n-2}{\csv} +  \f{\csv}  \cdot  \im{a}{2,n} \ef{n+2}{\csv} 
= \alp{1,n}{\csvb} \ef{n-2}{\csv} + \alp{2,n}{\csvb}  \ef{n+2}{\csv}.
\end{equation}
where we defined,
\begin{align}
\im{a}{1,n} &:= - \frac{\sqrt{n(n-1)}}{2}, ~~~  \im{a}{2,n} :=  \frac{\sqrt{(n+1)(n+2))}}{2}, \\
\alp{1,n}{\csvb} &:=  \f{\csv}  \cdot \im{a}{1,n}, ~~~ \alp{2,n}{\csvb} :=  \f{\csv}  \cdot \im{a}{2,n}.
\end{align}

\subsubsection{Structural Ingredients} \label{subsec:Space Adiabatic Perturbation Scheme}
Space adiabatic perturbation theory applies to the cosmological model, since the three conditions (see \ref{subsec:OM_Structural_Ingredients}), are satisfied. These are,
\begin{enumerate}
\item The quantum Hilbert space of the system decomposes as a tensor product,
\begin{equation}
\Hi = \Hs{\Lcs} \otimes \Hf{\Lcf},
\end{equation}
and the dynamics in $\Hs{\Lcs}$ happens on much larger scale as compared to the dynamics in $\Hf{\Lcf}$. In this model, the parameter $\pp:= \sqrt{\kappa/\lambda}$ represents the separation of these scales of change. 
\item
Deformation quantization with the Weyl-ordering can be employed for the cosmological model in order to make the space adiabatic perturbation scheme work on a technical level. 
\item The principal symbol of the Hamilton function, $\h{0}{\cscpb}$ possesses a pointwise isolated part of the spectrum $\im{\sigma}{0,\rel}^{\scriptscriptstyle{\cscpb}}$. We choose one of the eigenspaces with the energy label $\rel$. For distinct natural numbers $n \neq m; n, m \in \mathbb{N}$, the energy values of the corresponding bands never cross, 
\begin{align}
\left| \Efpd{n}{\oscpb}- \Efpd{m}{\oscpb} \right| = \olam\, \left| n- m \right| > 0.
\end{align}
\end{enumerate}
In the following, we work out the three steps for space adiabatic perturbation theory,
\begin{enumerate}
\item[1)] Construction of the Moyal projector $\pipd{\rel}{\cscpb} \in S^{\infty}(\pp; \Ps{\Lcs}, \mathcal{B}(\Hf{\Lcf}))$.
\item[2)] Construction of the Moyal unitary $\us{\rel}{\cscpb} \in S^{\infty}(\pp; \Ps{\Lcs}, \mathcal{L}(\Hf{\Lcf}))$.
\item[3)] Construction of the effective Hamiltonian $\heff{,\rel}{\cscpb}  \in S^{\infty}(\pp; \Ps{\Lcs}, \mathcal{L}(\Hf{\Lcf}))$.
\end{enumerate}

\subsubsection{Construction of the Moyal Projector $\pipd{n}{\cscpb}$}
For the motivation of the iterative construction of the Moyal projector, we refer the reader to section \ref{subsec:OM_Constr_pi}. In this regard, we define the $\csv$-dependent zeroth-order projectors on the eigenband associated to $\ef{\rel}{\csv} \in\Hf{\Lcf}$ as,
\begin{equation} \label{eq:pi0}
\pipd{0}{\csvb} := \ef{\rel}{\csv} \cdot \im{\left\langle \ef{\rel}{\csv}, \cdot \right\rangle}{\Lcf},
\end{equation}
where $\im{\left\langle \cdot, \cdot \right\rangle}{\Lcf}: \Hf{\Lcf} \times \Hf{\Lcf} \rightarrow \mathbb{C}$ denotes the inner product in $\Hf{\Lcf}$. The Moyal projector has the form of a formal power series in $\pp$,
\begin{equation} \label{eq:piformalseries}
\p = \sum_{N\geq0} \pp^N \pipd{N}{},~~~ \pipd{N}{} \in S^{\infty}(\Ps{\Lcs},\mathcal{B}(\Hf{\Lcf})),
\end{equation}
such that the actual Moyal projector is given as a resummation of \eqref{eq:piformalseries}. In analogy to the iterative construction prescription for the oscillator model in section\ref{subsec:OM_Constr_pi}, it is needed to compute the Moyal projector up to first order in perturbation theory $\pipd{(1)}{\cscp}$. Evaluating the first condition for $\pipd{1}{\cscp}$ and using the known Moyal projector at zeroth order, $\pipd{0}{\cscp}$, the determing equation for the diagonal part, $\pipd{1}{\text{D}}$ is,
\begin{equation}
\pipd{1}{} \cdot \pipd{0}{} + \pipd{0}{} \pipd{1}{} - \pipd{1}{} = 0.
\end{equation}
It implies, $\pipd{1}{\text{D}} = 0$. The third condition implies that the off-diagonal part, $\pipd{1}{\text{OD}}$, is determined by,
\begin{align} \label{eq:ProjectorSymbol1_1}
\pipd{1}{} = \frac{i}{2} &\left( (\h{0}{} - \Ef{\rel}\,\Uf{\Lcf})^{-1} (\Uf{\Lcf}
- \pipd{0}{}) \im{\left\lbrace \h{0}{} + \Ef{\rel}\,\Uf{\Lcf}, \pipd{0}{} \right\rbrace}{\Lcs} \pipd{0}{} \right. \nonumber \\
&~~\left.+\, \pipd{0}{} \im{\left\lbrace \pipd{0}{}, \h{0}{} + \Ef{\rel} \Uf{\Lcf} \right\rbrace}{\Lcs} (\h{0}{}
- \Ef{\rel}\,\Uf{\Lcf})^{-1} (\Uf{\Lcf}- \pipd{0}{}) \right). 
\end{align}
if $\Efpd{\rel}{\cscpb}=: \Ef{\rel}$ is a single, separated eigenvalue. Following the reasoning of section \ref{subsec:OM_Constr_pi}, we get for the Moyal projector at first order,
\begin{equation} \label{eq:pione_cos}
\pipd{1}{}\!=\!\left( \frac{\partial \Ef{\rel}}{\partial \csm} \right)\! 
\frac{i}{2 \im{\Delta}{E}}\! \left( \alp{1,\rel}{\csv} \left( \cprors{\rel}{-2}{\csvb} - \cprols{\rel}{-2}{\csvb} \right) 
+ \alp{2,\rel}{\csv} \left(  \cprols{\rel}{+2}{\csvb} - \cprors{\rel}{+2}{\csvb} \right) \right),
\end{equation}
where $\im{\Delta}{E} = \olam$ is now the minimal energy gap between two nearby energy functions of the Klein-Gordon subsystem. 

\subsubsection{Construction of the Moyal Unitary $\us{\rel}{\cscpb}$}
Analogously to the proceeding in section  \ref{subsec:Constr_MUnitary_OM}, we construct a unitary symbol $\us{}{} \in S^{\infty}(\pp; \Ps{\Lcs}, \mathcal{L}(\Hf{\Lcf}))$ which maps the dynamical subspace related to $\pipd{(1)}{}$, to a suitable reference subspace $\im{\Hi}{0}$.
This unitary symbol is associated to the eigensolution $\ef{\rel}{\csvz}$ for some fixed $\csv \equiv \csvz$, i.e.,
\begin{equation}
\pipd{\refs}{} := \ef{\rel}{\csvz} \im{\left\langle  \ef{\rel}{\csvz}, \cdot \right\rangle}{\Lcf}. 
\end{equation}
The unitary operator, which is in line with the conditions \ref{itm:1} and \ref{itm:3} at zeroth order in section \ref{subsec:Constr_MUnitary_OM}, is given by,
\begin{equation}
\us{0}{\csvb} = \sum_{j \geq 0} \ef{j}{\csvz} \im{\left\langle \ef{j}{\csv}, \cdot \right\rangle}{\Lcf}. 
\end{equation}
Then, taking $\us{0}{}$ as a starting point, and referring to the iterative construction conditions again, it is straightforward to show that the hermitian 
part of $\us{1}{}$ vanishes because $\us{0}{}$ is independent of $\csm$. However, the remaining 
anti-hermitian contribution is given according to \eqref{eq:Unitary_Firstorder_OM}, by
\begin{equation}
\us{1}{}\!=\!\left(\frac{\partial \Ef{\rel}}{\partial \csm} \right) \frac{i}{2 \im{\Delta}{E}}\!\left(\alp{1,\rel}{\csvb} \left(  \cproru{\rel}{-2}{(\csvz)}{\csvb} + \cprolu{\rel}{-2}{(\csvz)}{\csvb} \right) 
- \alp{2,\rel}{\csvb} \left( \cproru{\rel}{+2}{(\csvz)}{\csvb} + \cprolu{\rel}{+2}{(\csvz)}{\csvb} \right) \right).
\end{equation}

\subsubsection{Construction of the Effective Hamilton Symbol $\heff{,\rel}{\cscpb}$}
The last step of the perturbation scheme consists in computing the effective Hamiltonian $\heff{}{\cscpb}$ up to second order, which is 
assumed to have the form of a formal power series,
\begin{equation}
\heff{,(2)}{\cscpb} = \sum_{N = 0}^2 \pp^N\, \heff{,N}{\cscpb},~~ \heff{,N}{\cscpb} \in S^{\infty}(\Ps{\Lcs}, \mathcal{L}(\Hf{\Lcf})).
\end{equation}
The defining equation for $\heff{,(2)}{\cscpb}$ at $N$-th order is given by \eqref{eq:OM_heffCondk}. The focus of the computations lies primarily in the dynamics within the subspace related to the fast quantum number $\rel$. The related effective Hamilton symbol, $\heff{,\rel,(2)}{} := \pipd{\refs}{} \cdot \heff{,\rel,(2)}{} \cdot \pipd{\refs}{}$ at zeroth order is given by,
\begin{equation} \label{eq:ZeroHeff}
\heff{,\rel,0}{\cscpb}
= \left( -\frac{\csm^2}{2 \cmg} + \frac{1}{2} \cmg \ok^2 \csv^2  
+ \olam  \left( \rel + \frac{1}{2} \right) \right) \pipd{\refs}{}.
\end{equation}
Thus, the effective Hamilton symbol for the gravitational degrees of freedom at zeroth order includes the bare gravitational Hamilton symbol plus an off-set energy which stems from the Klein-Gordon's particle chosen energy value. This result equals the outcome of the Born-Oppenheimer approximation for this model.\\
As in the oscillator model, the first order contribution of the effective Hamilton symbol, $\heff{,1}{\cscpb}$ contains only off-diagonal terms, such that $\heff{,\rel,1}{\cscpb}$ vanishes, 
\begin{equation}
\heff{,\rel,1}{\cscpb} = \frac{i}{2}\, \pipd{\refs}{} \im{\left\lbrace \us{0}{}, \h{0}{}
+ \Ef{\rel}\,\Uf{\Lcf} \right\rbrace}{\Lcs}\, (\us{0}{})^{\ast}\, \pipd{\refs}{} =0.
\end{equation}
The reasoning for  $\heff{,\rel,1}{\cscpb}$ gives for the second order contribution the result analogous to the result in \eqref{eq:OM_heff_second}. The explicit symbol for the cosmological model is thus,
\begin{equation} \label{second}
\heff{,\rel,2}{\cscpb} =  \frac{1}{2\,\cmg \olam} \left( - \frac{\csm^2}{\ok \csv^2} \left(\rel 
+ \frac{1}{2} \right) - \frac{1}{\csv^2} \frac{\olam}{2} \left(\rel^2 + \rel+1 \right)\right) \pipd{\refs}{}.
\end{equation}
This proves our statement that besides the trivial Born-Oppenheimer approximation, further backreaction effects arise for the gravitational subsystem. It is now easy to evaluate the action of this symbol on some generic tensor product wave function in $\Hi = \Hs{\Lcs} \otimes \Hf{0}$, since the Klein-Gordon tensor factor does not depend on the gravitational degrees of freedom anymore.

\subsubsection{Eigensolutions of the Effective Hamilton Operator} \label{sec:Eigensolutions}

The final step consists in the Weyl quantisation of the effective Hamiltonian 
symbols constructed above up to second order in the adiabatic parameter which consists of 
the two contributions \eqref{eq:ZeroHeff} and \eqref{second} of zeroth and second 
order respectively. The zeroth order contribution has no ordering ambiguities and represents essentially an inverted harmonic oscillator with non-vanishing
zero point energy. The second order contribution gets symmetrically Weyl-ordered in this step and involves negative powers of $\csv$, thus is more singular than the zeroth order contribution.

Considering zeroth order contribution first, the spectral problem of the inverted 
oscillator is well studied in the literature, see e.g. \cite{21} and references therein.
The spectrum is of the absolutely continuous type and the operator is essentially self-adjoint on the space of smooth functions vanishing at infinity.
The generalised (i.e. not normalisable) eigenvectors are explicitly known in terms of parabolic cylinder functions. It is possible to choose boundary conditions such that these eigenfunctions vanish at zero where the classical singularity resides. 

One would now like to proceed as for the oscillator model and treat the second order contribution as a
small correction to the second order by means of stationary perturbation theory. Unfortunately, this method is applicable only when the zeroth order has a pure point spectrum. In fact, it is well known that the perturbation theory for absolutely continuous operators is very unstable in the sense that a perturbation by an operator of arbitrarily small Hilbert-Schmidt norm exists such that their 
sum has pure point spectrum \cite{20}. We are not aware of any rigorous work in that direction and it seems that the spectral problem of the Hamiltonian 
constraint operator including zeroth and second order contributions cannot use simple perturbative methods but must be addressed by independent methods such as using the dense and invariant domain studied in \cite{22}.

We thus leave this problem for future research and just point out once more how non-trivial the inclusion of backreaction effects can become.

\section{Conclusion and Outlook}
\label{s4}

We have applied the SAPT machinery to two closely related quantum mechanical models 
which capture aspects of the purely homogeneous sector of quantum cosmology in order 
to study the influence of backreaction between geometry and matter contributions 
to the Hamiltonian constraint. This influence can be encoded in terms of 
a purely gravitational effective Hamiltonian which receives corrections from the 
backreactions that one would not expect in a crude Born-Oppenheimer approximation.
These correction terms could play an important role in the details of a possible 
singularity resolution that was discovered in the LQC approach\cite{9}. While 
we have not done so in this paper, one could of course also quantise the slow 
sector using LQC methods, one would only need to adapt the Weyl quantisation 
map from the phase space $T^\ast (\mathbb{R})$ to $T^\ast (S^1)$ which was already
done in \cite{14}. We have not done so, because \cite{22} shows that there exists 
a dense and invariant domain for the Hamiltonian constraint and all its 
finite order backreaction terms, in particular all matrix elements can be computed 
in closed form. We plan to come back to the phenomenological consequences of 
backreactions for the model discussed in this paper in future publications.

In the two remaining papers of this series we will leave the relatively safe realm
of quantum mechanics and enter quantum field theory by taking the inhomogeneous 
degrees of freedom into account to linear order in cosmological (rather than 
adiabatic) perturbation theory \cite{17,18}. There we will face additional challenges 
that were already explained in \cite{16}: The SAPT scheme is not immediately in the 
most natural variables that one would expect which has to do with unitary equivalence of
Fock representations (or not). One can solve these problems using adapted variables 
the technique for which was found in \cite{15} but this can lead to {\it tachyonic quantum 
fields} in some parts of the slow (homogeneous) phase space. An example for a  
(gauge invariant) quantum field with indefinite 
mass squared term is the Mukhanov-Sasaki scalar perturbation in cosmology. 
In \cite{16} we have made several proposals for overcoming the corresponding problems 
including a phase space restriction induced by an additional partial 
canonical transformation in the homogeneous sector and will apply them in \cite{17,18}.\\
\\
~\\
{\bf Acknowledgements}\\
\\
J.N. thanks the Max Weber Stiftung for financial support. S.S. thanks the 
Heinrich-B\"oll Stiftung for financial and intellectual support and the 
German National Merit Foundation for intellectual support.

}

\end{document}